\documentclass[aps,prl,preprint,groupedaddress]{revtex4}
\usepackage{graphicx}
\input{epsf}

\begin{document}

\title{Bias dependent features in spin transport as a\\
probe of the conduction band minima in Si} 

\author{S. Parui, K. G. Rana and T. Banerjee}
\affiliation{Physics of Nanodevices, Zernike Institute for Advanced Materials, University of Groningen, Nijenborgh 4, 9747 AG Groningen, The Netherlands}

\date{\today}

\begin{abstract}
Unusual features in the bias dependence of spin transport are observed in a Co/Au/NiFe spin valve fabricated on a highly textured Cu(100)/Si(100) Schottky interface, exploiting the local probing capabilities of a Ballistic electron magnetic microscope (BEMM). This arises due to local differences in the strain and the presence of misfit dislocations at the Schottky interface that enhances spin flip scattering and broadens the energy and angular distribution of the transmitted electrons. Cumulatively, these enable the transmitted hot electrons to probe the different conduction band minima in Si, giving rise to such bias dependent features in the magnetocurrent. This study reveals new insights into the spin dependence of transmission in an indirect band gap semiconductor as Si and highlights the unique capabilities of BEMM in probing local differences in spin transport across such textured interfaces.

\vspace{1pc}
PACS numbers: {72.25.Ba, 75.47.-m,  85.75.-d}
\vspace{1pc}
\end{abstract}

\maketitle

Metallic spin valves on Schottky interfaces are ideal for investigating spin transport in ferromagnetic (FM) layers, studying spin asymmetry of relaxation times for electrons and holes in FMs, probing spin-sensitivity of transmission at the electronically dissimilar FM-Semiconductor (SC) heterointerface, etc. This is exemplified by numerous studies involving hot electron transport in metallic spin valves on different semiconductors as Si \cite{Monsma, Ian, Rippard, Tamalika1, Tamalika2}, GaAs \cite{Parkin}, GaP \cite{Reddy}, GaAsP \cite{Heindl} etc. Probing a different regime of the electronic band structure, such studies involving hot electrons and holes have yielded quantitative estimates of different transport parameters as the scattering time and attenuation length at energies a few eV above and below E$_F$, \cite{TamalikaPRL} and have been utilized for designing new schemes and devices in spintronics. Studies on spin transport reported so far have been performed using spin valves with different FMs as NiFe, Co, Fe, Ni, etc. on  polycrystalline Schottky interfaces with Si. In GaAs, a direct band gap semiconductor, epitaxial Schottky interfaces are formed with Fe or its alloys and spin dependent transmission has revealed nonmonotonic bias dependence of the magnetocurrent arising from the different conduction band structures in GaAs \cite{Parkin}.\\

Here, we demonstrate an unique approach of exploiting hot electron spin transport in a Co/Au/NiFe spin valve grown on an epitaxial oriented Schottky interface of Cu(100)/Si(100), that reveal signatures of the electronic structure of an indirect band gap semiconductor of Si (100). This is demonstrated by analyzing the collected current, on the nanoscale, at different regions of the textured Schottky interface using the local probing capabilities of the Ballistic electron magnetic microscope (BEMM). Although it is known that ultrathin Cu grows epitaxially on Si(100), \cite{Cugrowth, Parui} such interfaces have not been exploited earlier to study spin dependence of transmission in metallic spin valves. Interestingly, for different locations in the same spin valve device on Si(100) (denoted as Regions 1 and 2), we find that the collected current for both parallel (I$_P$) and antiparallel (I$_{AP}$) alignments of the magnetic layers exhibit significant differences in their bias dependence. The overall transmission for I$_P$, in Region 2 (R2) is generally found to be lower  than in Region 1 (R1) upto a bias $\sim$ -1.4 V and thereafter increases gradually. For the antiparallel case, the increase is more abrupt at a similar bias. Further, the collected current in R2 exhibits unusual features in its bias dependence beyond $\sim$ -1.4 V. A large magnetocurrent (MC) of $\sim$600$\%$ (at -0.9 V) [MC = (I$^P$ - I$^{AP}$)/I$^{AP}$] is predominantly observed (in R1) which decreases to $\sim$200$\%$  (at -2 V) with a monotonic bias dependence. The extracted magnetocurrent in R2 also decreases and reaches $\sim$100$\%$  at -2 V.\\   

To study the bias dependence of spin transport, we have used Ballistic electron magnetic microscopy (BEMM). BEMM is a three electrode extension (Fig. 1) of the Scanning tunneling microscopy (STM) that is used to investigate perpendicular spin transport through buried layers and interfaces in a spin valve grown on a semiconducting substrate \cite{Rippard, Tamalika1}. A PtIr STM tip is used to locally inject electrons onto the top Au layer and an additional contact at the semiconductor rear serves as the collector \cite{Bell}. In BEMM, the energy of the injected electrons can be varied by changing the applied bias V$_T$ for a fixed tunnel current I$_t$, thus yielding fundamental insights into the energy dependence of the scattering processes in the ferromagnetic layers. The spin dependence of the collected current, I$_B$, is obtained by applying a magnetic field that changes the relative magnetization alignment of the FM layers in the spin valve. After transmission through the spin valve, the electrons are collected in the conduction band of the Si semiconductor, provided they satisfy the necessary energy and momentum criteria at the M/S Schottky interface. The spin dependent collector current, I$_B$, for the parallel and antiparallel configuration, can be written as,

\begin{equation}
I^P_B\propto(T^M_{FM1}T_{S}T^M_{FM2}+T^m_{FM1}T_{S}T^m_{FM2})\Upsilon^P_{Cu/Si}
\end{equation}
\begin{equation}
I^{AP}_B\propto(T^M_{FM1}T_{S}T^m_{FM2}+T^m_{FM1}T_{S}T^M_{FM2})\Upsilon^{AP}_{Cu/Si}
\end{equation}

where T$^M$ and T$^m$ refer to the transmission of the majority (M) and minority (m) hot electrons in the ferromagnetic layers, and T$_S$ is the transmission in the spacer (non-magnetic; NM) layer. The bulk transmission depends exponentially on the NM and FM layer thickness as, $T\propto e^{-d/\lambda}$, where $d$ is the thickness and $\lambda$ the hot electron attenuation length for the NM and FM layers [see inset of Fig. 1]. $\Upsilon$$_{Cu/Si}$ denotes the scattering sensitive transmission probability at the highly textured M/S interface.\\ 

For the BEMM experiments, we use n-type Si(100) substrate with a 300 nm thick SiO$_2$ which is removed by buffered hydrofluoric acid (HF). The pre-defined circular devices, 150 $\mu$m in diameter, are hydrogen terminated using 1$\%$ hydrofluoric acid, onto which metal layers were deposited by e-beam evaporation. First, a 10 nm Cu layer was evaporated to form a highly textured Cu/Si Schottky barrier \cite{Cugrowth}, followed by the deposition of 4 nm Co, 10 nm of Au and 4 nm of NiFe films. A final 5 nm of Au film serves as a capping layer, for \textit{ex situ} sample transfer to the BEMM set up.\\

For an unambiguous demonstration of spin transport in such spin valves, nanoscale magnetic hysteresis loops were recorded using BEMM as shown in Fig. 2. The magnetic field dependence of the transmission, measured at -2 V and 4 nA tunnel injection shows clear switching, characteristic of the relative parallel and antiparallel alignment of the NiFe and Co layers. A typical measurement starts by fully saturating and aligning the magnetization of both the FMs by ramping the magnetic field to a maximum. When the magnetic field is swept through zero and changes sign, the softer FM viz. NiFe switches leading to an anti-parallel (AP) alignment and reduction in BEMM current. A further increase in the magnetic field switches the Co layer with the magnetizations aligned parallel (P) resulting in a higher BEMM current. The inset confirms independent switchings of the NiFe and Co layers in the spin valve, using SQUID (Superconducting Quantum Interference Device) and characterized by a stepped hysteresis loop with two distinct coercive fields, \cite{Lucinski} matching well with that of the magnetic hysteresis loop obtained using BEMM.\\

Spin-dependent transmission as measured in BEMM, for both parallel and anti-parallel magnetization orientations of the Co(4)/Au(10)/NiFe(4)/Au(5) spin valve are shown in Figs. 3 a. and b. Figure 3 a. represents the transmission as obtained in approximately 80 $\%$ of the device area in Si/Cu(10)/Co(4)/Au(10)/NiFe(4)/Au(5) and designated as R1. However, by moving the STM tip across the device, the BEMM transmission as well as the spectral shape is found to differ (as shown in Fig. 3 b.), at few other locations (approximately 20 $\%$ of the device area), from that in R1, and denoted here as R2. At such locations, we also observe unusual features in the bias dependence for both the I$_P$ and I$_{AP}$ transmissions. The overall transmission for I$_P$ is lower than that in R1 upto a bias $\sim$ -1.4 V and thereafter increases gradually. For the antiparallel case, the increase is more abrupt at a similar bias. The spectral shape of the collected current for the parallel and antiparallel transmission reflects the narrow energy distribution of the transmitted majority electrons and a relatively broad one for the minority electrons.  The extracted magnetocurrent for the two regions is shown in Fig. 3 c. The MC decreases with bias for both the regions but is more abrupt for R2 and reaches $\sim$100$\%$ at -2 V. To rule out possible artifacts in the BEMM transmission related to the growth of Co on Cu, we have also fabricated and measured spin dependent tranport in a Si/Cu(10)/NiFe(4)/Au(10)/Co(4)/Au(5) device (i.e the bottom FM is now NiFe). Here too (Supplementary Figure S1), we find the occurrence of two regions with a similar transmission characteristics and bias dependence of MC (Fig. S1).\\

The non trivial bias dependence of the BEMM transmission is further analysed in Fig. 4. It shows the data of the BEMM transmission for P and AP configuration, in Regions 1 and 2, normalized to its value at -2 V (left) and -1.2 V (right).  We find that the spectral shape and the energy dependence of the BEMM transmission in R2 is different than in R1, for both the P and AP alignments, when normalized at -2 V. Normalizing the transmissions at a lower energy allows us to identify the onset in the bias beyond which features in the P and AP spectra sets in. This has been done at several lower biases and is shown for a bias of -1.2 V (Fig. 4 right). We see that beyond $\sim$ -1.4 V, the energy dependence of both the P and AP transmission becomes more pronounced and correspond to the energy beyond which the unusual features in the BEMM transmission sets in.\\

The spin dependence of transmission of the injected electrons propagating through the spin valve stack (T$^{M,m}$$_{FM1,2}$) and their interfaces are expected to be similar in both Regions 1 and 2 of the device. However, inhomogeneity at the M/S interface in the device, due to differences in the strained epitaxial lattice and/or the presence of misfit dislocations acts as effective elastic scattering centres, as in R2, and increases the transmission probability ($\Upsilon$$_{Cu/Si}$) of the transmitted electrons, above $\sim$ -1.4 V, at such M/S interfaces. These scattering events also broaden the energy and angular dependence of the transmitted electrons and increases the relative fraction of minority electrons by spin flip processes at the interface (increase in I$_{AP}$). This is accompanied by a concomitant reduction in the spin asymmetry (I$_P$/I$_{AP}$) at higher energies (above $\sim$ -1.4 V) and is manifested as an abrupt decrease in the MC with bias in R2. Subsequently, this energy and momentum broadening of the transmitted electrons enables the electrons with larger parallel wave vectors to access the almost degenerate minima of the second conduction band in addition to the first conduction band at the X point (85$\%$ from the $\Gamma$ point) in Si, besides probing the second minimum in the first conduction band at the L point as shown in Fig. 5. The unusual features observed in the bias dependence of spin transport in R2 corresponds well to these additional thresholds in the conduction band minima in Si at both the X (corresponds to collection from and above the Schottky barrier height of 0.62 V at the Cu/Si interface) and L point (corresponds to $\sim$ 0.8 V above the Schottky barrier height) \cite{Cohen, Klein}. Thus, the observed effect at $\sim$ -1.4 V is 0.8 V above the Schottky barrier height at the M/S interface and matches with the energy separation between the first conduction band minimum at X and the second conduction band minimum at L in Si. \\

This possibility to probe the additional conduction band minima in an indirect band gap semiconductor as Si was not demonstrated in earlier studies involving polycrystalline M/S interfaces. It is enabled by the unique choice of an epitaxial M/S interface and the capabilities of the BEMM in resolving local differences in spin transport, in the same device, that arises due to differences in the local strain at the underlying epitaxial oriented Cu/Si interface. The local differences in the epitaxial oriented Cu/Si interface due to misfit dislocations in R2, enhances momentum scattering and broadens the energy and angular distribution of the transmitted electrons. This in turn increases the minority electrons due to spin flip processes, decreases the transmission asymmetry at higher energies and reduces the MC. All these cumulatively enables the hot electrons with larger parallel wave vectors to access the X and L bands in an indirect band gap material as Si giving rise to such features in their spin transport, hitherto not reported, and is a new approach to probe the electronic band structure in semiconductors.\\

We thank A. M. Kamerbeek for scientific discussions, T. T. M. Palstra and B. Noheda for use of the XRD and  MPMS. Technical  support  from  J.  Holstein,  J.  Baas,  B.  Wolfs  and  M.  de  Roosz  is thankfully acknowledged. This work was financially supported by the NWO-VIDI program, the Zernike Institute for Advanced Materials and NanoNed program coordinated by the Dutch Ministry of Economic Affairs.

\clearpage

\begin{figure}
\includegraphics[scale=0.65]{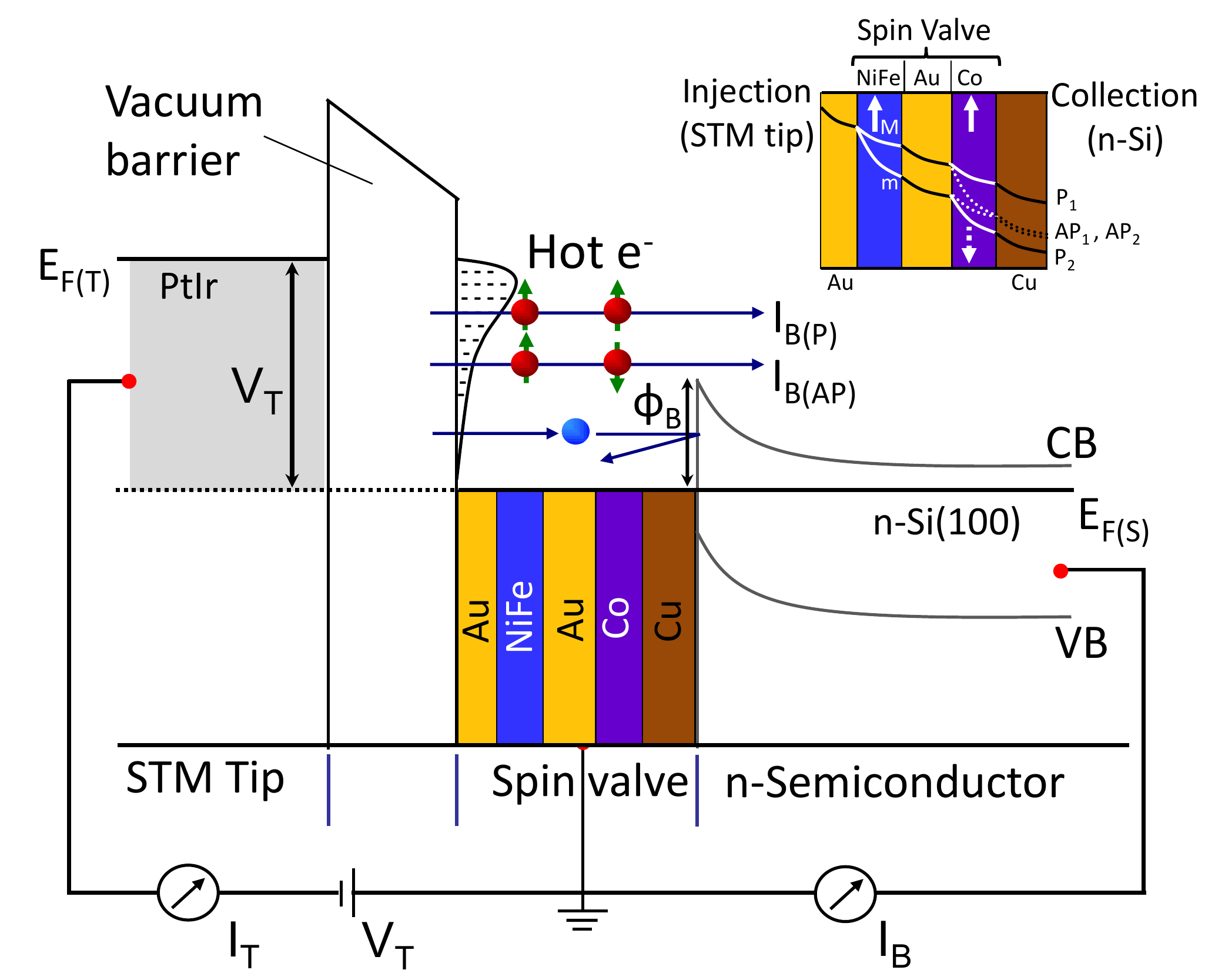}
\caption{\label{fig:BEMM schematic} Schematic energy diagram of the BEMM experiment. The sample consists of a n-Si(100) substrate coated with Cu(10)/Co(4)/Au(10)/NiFe(4)/Au(5) (nm). The STM tip is used to locally inject hot electrons into the multilayer base by tunneling at a bias voltage V$_T$ between tip and Au surface. The current I$_B$ flows perpendicular through the stack and is collected at the conduction band of the Si substrate. (Top right) shows spin-dependent transmission of the majority and minority electrons at the normal metal/ferromagnetic metal interface and their exponential decay in the bulk of the metal films. The P and AP transmission are described in Equation 1 and 2.}
\end{figure}

\begin{figure}
\includegraphics[scale=0.60]{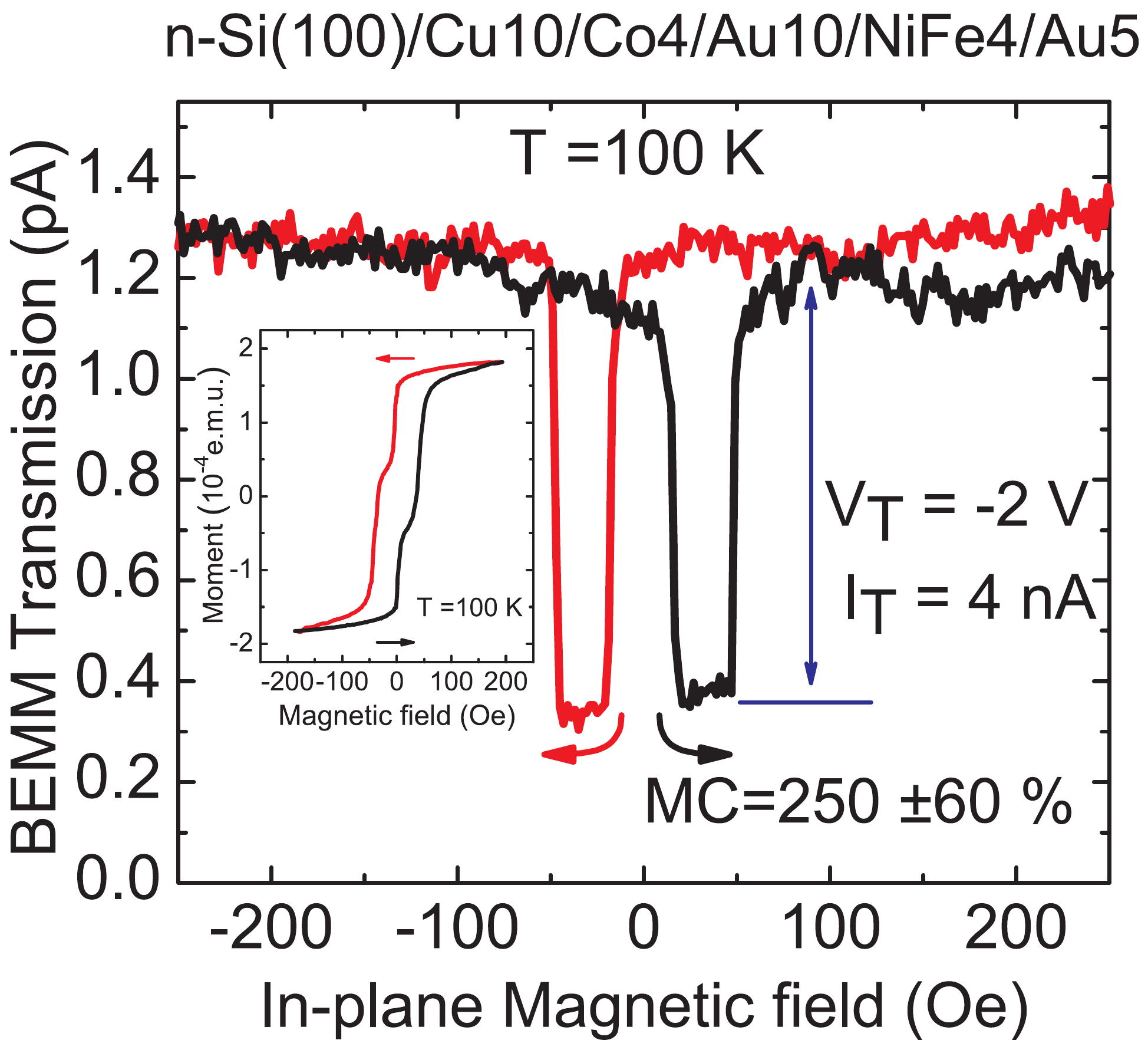}
\caption{\label{fig:SQUID} a and b: Local hysteresis loop obtained by BEMM at tunneling voltage of -2 V and at tunneling current of 4 nA, showing clear P and AP transmissions. NiFe switches at $\sim$16 Oe and the switching field of Co on Cu is $\sim$50 Oe. These values are consistent with that obtained from SQUID measurement. (Inset) SQUID magnetometer measurement, showing switching fields of NiFe and Co. Arrows indicate magnetic field sweep direction. Measurements exhibit hysteresis loops with well-defined plateaus representing anti-parallel configuration of Co and NiFe magnetic moments.}
\end{figure}

\begin{figure}
\includegraphics[scale=0.62]{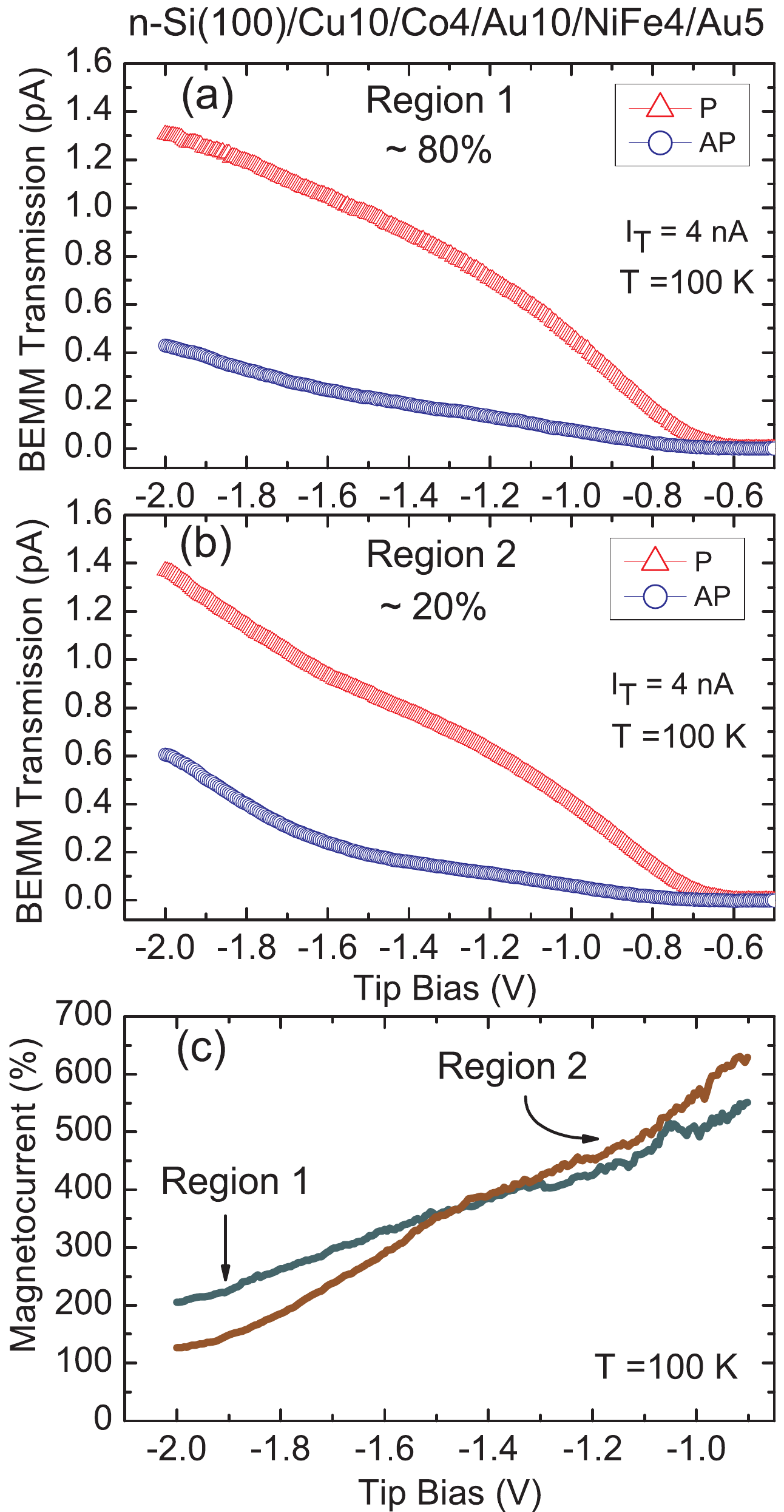}
\caption{\label{fig:SQUID} a, b: The variation of the BEMM transmissions for parallel (P) and anti-parallel (AP) magnetization configuration of the Si/Cu(10)/Co(4)/Au(10)/NiFe(4)/Au(5) device corresponding to two different regions in the same device. For both regions the transmission in the P region is larger than that at AP. In region 2, a second onset is found at $\sim$ -1.2 V for both P and AP transmission unlike that in Region 1. c: From the parallel and antiparallel BEMM transmission, magnetocurrent is extracted for both Regions 1 and 2 and plotted with respect to the bias.}
\end{figure}

\begin{figure}
\includegraphics[scale=0.68]{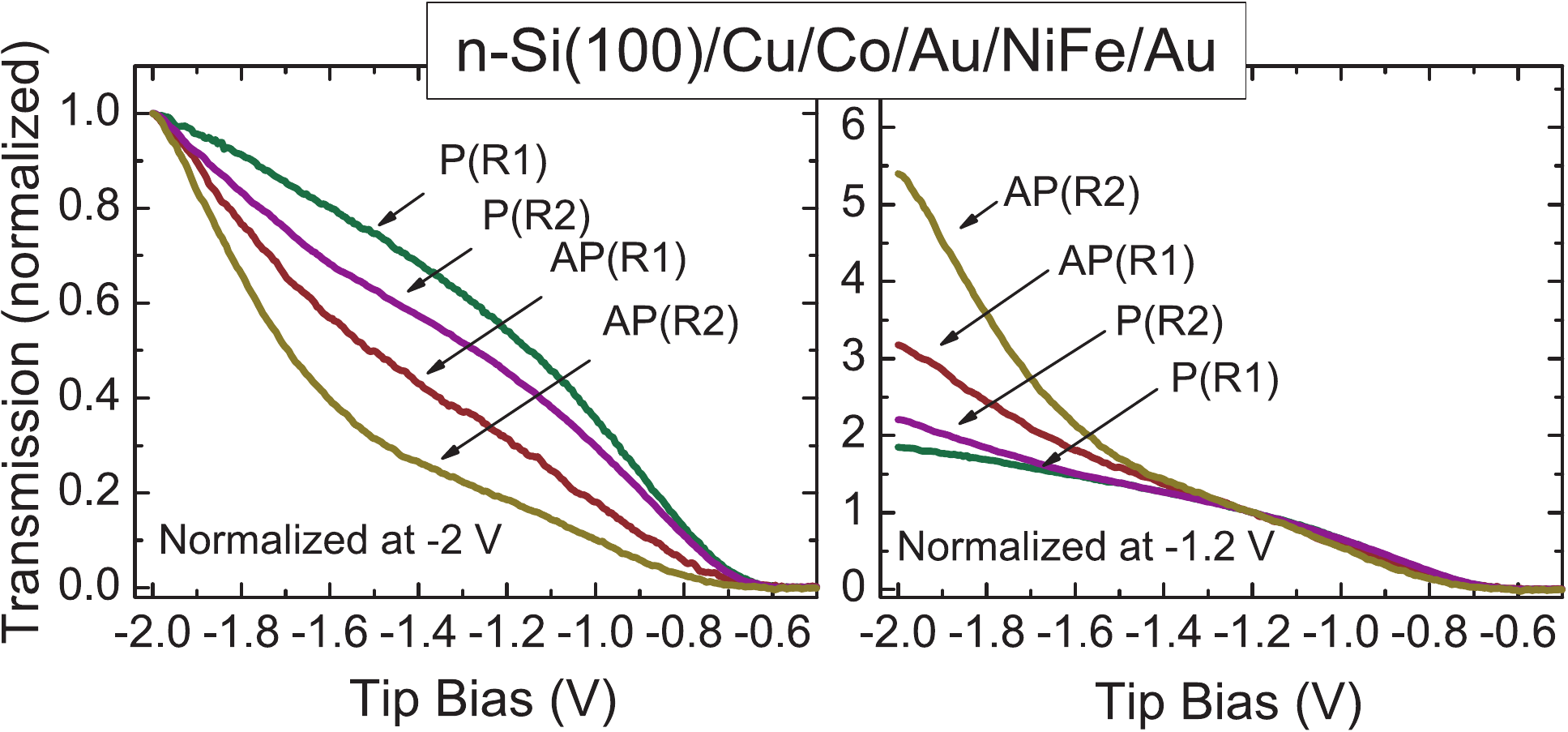}
\caption{\label{fig:SQUID} Normalized transmissions for the Si/Cu(10)/NiFe(4)/Au(10)/Co(4)/Au(5) device from Fig. 3. Both the P and AP transmissions are normalized at tip biases of -2 V  (left) and -1.2 V (right) respectively. Normalized plots at -2 V show a distinct change in the direction of curvatures indicating different energy dependence for Region 2. When the BEMM transmissions are normalized at -1.2 V, the onset of the energy dependence of the transmissions becomes more pronounced, as shown.}
\end{figure}

\clearpage 

\begin{figure}
\includegraphics[scale=0.78]{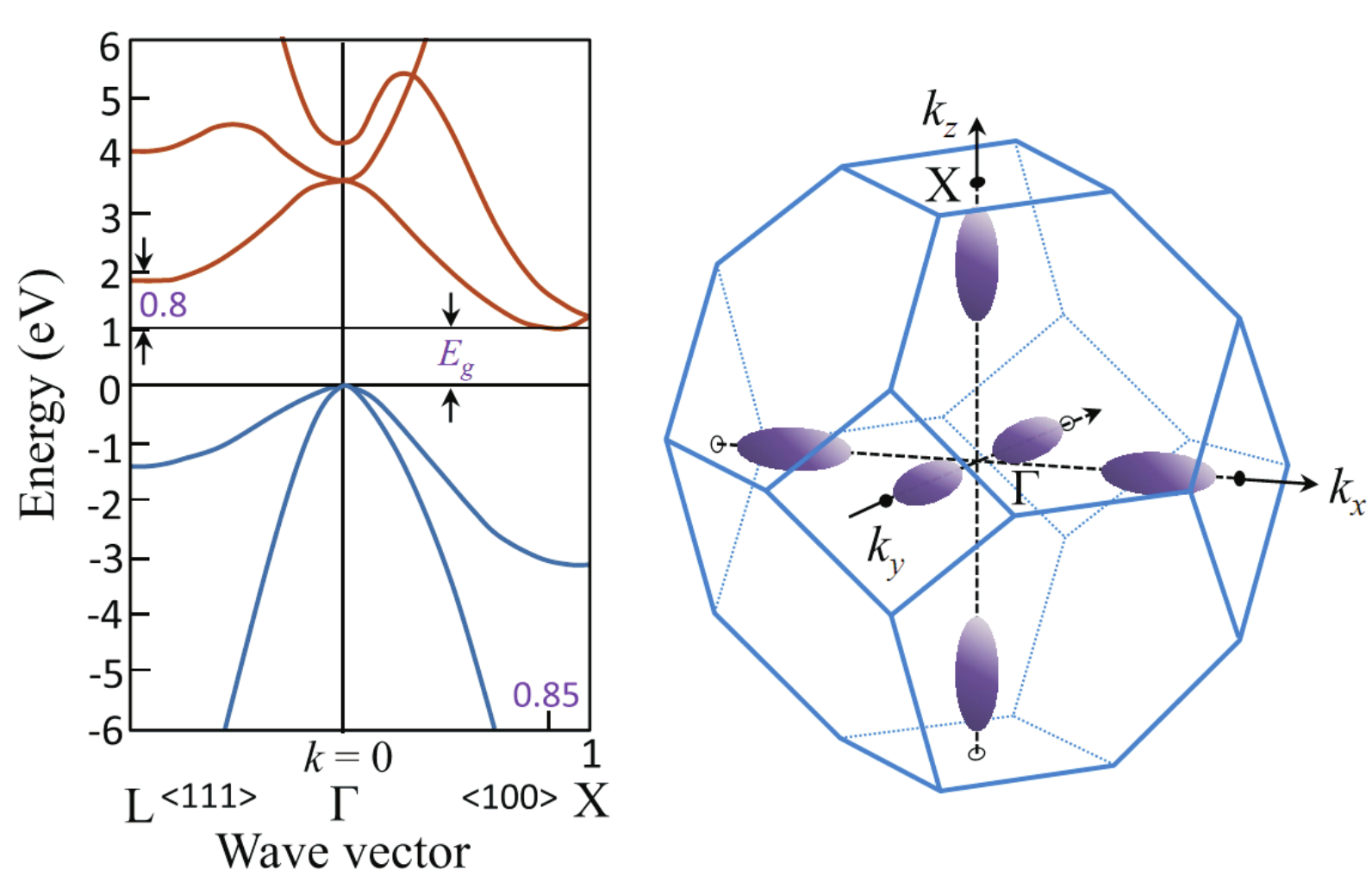}
\caption{\label{fig:SQUID} 
Electronic structure of Si. There are six equivalent ellipsoidal conduction band minima at a distance about 85$\%$ from the $\Gamma$ point to the X points.  The second conduction band minima is only 0.1 eV above the minima of the first conduction band at the X-points. The second minima of the first conduction band is at the L point which is about 0.8 eV above the first minima.}
\end{figure}


\vspace{4pc}


\begin{thebibliography}:
\bibitem{Monsma} D. J. Monsma, R. Vlutters, and J. C. Lodder, Science {\bf 281}, 407 (1998); D. J. Monsma, J. C. Lodder, Th. J. A. Popma, and B. Dieny Phys. Rev. Lett. {\bf 74}, 5260 (1995).
\bibitem{Ian} I. Appelbaum, B. Huang, and D. J. Monsma, Nature (London) {\bf 447}, 295 (2007); B. Huang, D. J. Monsma and I. Appelbaum, Phys. Rev. Lett. {\bf 99}, 177209 (2007).
\bibitem{Rippard} W. H. Rippard and R. A. Buhrman, Appl. Phys. Lett. {\bf 75}, 1001 (1999); Phys. Rev. Lett. {\bf 84}, 971 (2000).
\bibitem{Tamalika1} T. Banerjee, J. C. Lodder, and R. Jansen, Phys. Rev. B {\bf 76}, 140407(R) (2007).
\bibitem{Tamalika2} T. Banerjee, W. G. van der Wiel, and R. Jansen, Phys. Rev. B {\bf 81}, 214409 (2010).
\bibitem{Parkin} S. van Dijken, X. Jiang, and S. S. P. Parkin, Phys. Rev. Lett. {\bf 90}, 197203 (2003); X. Jiang, S. van Dijken, R. Wang and S. S. P. Parkin, Phys. Rev. B. {\bf 69}, 014413 (2004).
\bibitem{Reddy} C. V. Reddy, R. E. Martinez II, V. Narayanamurti, H. P. Xin, and C. W. Tu, Phys. Rev. B {\bf 66}, 235313 (2002).
\bibitem{Heindl} E. Heindel, J. Vancea, and C. H. Back, Phys. Rev. B {\bf 75}, 073307 (2007).
\bibitem{TamalikaPRL} T. Banerjee, E. Haq, M. H. Siekman, J. C. Lodder and R. Jansen, Phys. Rev. Lett. {\bf 94}, 027204 (2005).
\bibitem{Cugrowth} H. Jiang, T. J. Klemmer, J. A. Barnard, and E. A. Payzant, J. Vac. Sci. Technol. A {\bf 16}, 3376 (1998); B. G. Demczyk, R. Naik, G. Auner, C. Kota, and U. Rao, J. Appl. Phys. {\bf 75}, 1956 (1994).
\bibitem{Parui} S. Parui, J. R. R. van der Ploeg, K. G. Rana and T. Banerjee, Phys. Status Solidi RRL {\bf 5}, 388 (2011).
\bibitem{Bell} W. J. Kaiser and L. D. Bell, Phys. Rev. Lett. {\bf 60}, 1406 (1988); L. D. Bell and W. J. Kaiser, Phys. Rev. Lett. {\bf 61}, 2368 (1988).
\bibitem{Lucinski} T. Luci\'{n}ski, S. Czerkas, H. Br\"{u}ckl, and G. Reiss, J. Magn. Magn. Mater. {\bf 222}, 327 (2000).
\bibitem{Cohen} M. L. Cohen and T. K. Bergstresser, Phys. Rev. {\bf 141}, 789 (1966).
\bibitem{Klein} C. S. Wang and B. M. Klein, Phys. Rev. B {\bf 24}, 3393 (1981).


\end{thebibliography}
\end{document}